\documentstyle[12pt,aasms4]{article}
\newcommand{\ms}{\mbox{m s$^{-1}~$}}
\newcommand{\msun}{M$_{\odot}$}

\newcommand{\mjup}{M$_{\rm JUP}$}

\newcommand{\msini}{$M \sin i~$}

\lefthead{Marcy {\it et~al.\/}}
\righthead{Planetary Companion to Gliese 876}
\slugcomment{Submitted to Astrophysical Journal Letters}
\received{}
\accepted{}
\begin{document}

\title{A Planetary Companion to the Nearby M4 Dwarf, Gliese 876$~^{1}$}

\author{Geoffrey W. Marcy\altaffilmark{2},
R. Paul Butler,\altaffilmark{3}, 
Steven S. Vogt\altaffilmark{4}, 
Debra Fischer\altaffilmark{2},
Jack J. Lissauer\altaffilmark{5}}

\authoremail{gmarcy@etoile.berkeley.edu}

\altaffiltext{1}{Based on observations obtained at Lick Observatory, which
is operated by the University of California, and on observations obtained
at the W.M. Keck Observatory, which is operated jointly by the
University of California and the California Institute of Technology.}

\altaffiltext{2}{Department of Physics and Astronomy, San Francisco, CA, USA 94132
and at Department of Astronomy, University of California,
Berkeley, CA USA  94720}

\altaffiltext{3}{Anglo--Australian Observatory, PO Box 296, NSW 2121 Epping, Australia}

\altaffiltext{4}{UCO/Lick Observatory, 
University of California at Santa Cruz, Santa Cruz, CA, 95064}

\altaffiltext{5}{NASA/Ames Research Center, 
NASA/Ames Research Center, MS245-3, Moffett Field, CA 94035}

\begin{abstract}

Doppler measurements of the M4 dwarf star, Gliese 876, taken at both
Lick and Keck Observatory reveal periodic, Keplerian velocity
variations with a period of 61 days.  The orbital fit implies that the
companion has a mass of, $M$ = 2.1 M$_{\rm JUP}/ \sin i$, an orbital
eccentricity of, $e$ = 0.27$\pm$0.03, and a semimajor axis of, $a$ =
0.21 AU.  The planet is the first found around an M dwarf, and was
drawn from a survey of 24 such stars at Lick Observatory.  It is the
closest extrasolar planet yet found, providing opportunities for
follow--up detection.  The presence of a giant planet on a
non-circular orbit, 0.2 AU from a 1/3 M$_{\odot}$ star, presents a
challenge to planet formation theory.  This planet detection around an
M dwarf suggests that giant planets are numerous in the Galaxy.
\end{abstract}

\keywords{planetary systems -- stars: individual (Gliese 876)}

\section{Introduction}
\label{intro}

Precise Doppler surveys of main sequence stars have revealed eight
companions that have masses under 5 \mjup / $\sin i$, with the orbital
inclination, $i$, remaining unknown (Mayor et al. 1999, Marcy \&
Butler 1998, Noyes et al. 1997, Cochran et al. 1997).  These
``planetary'' companions exhibit both circular and eccentric orbits,
consistent with formation in dissipative circumstellar disks, followed
by gravitational perturbations (cf. Lin et al. 1995, Artymowicz 1997,
Levison et al. 1998).  The semimajor axes are all less than 2.5 AU,
with most being less than 0.3 AU.  This ``piling--up'' of planets near
their host stars appears to be a real effect, although enhanced by the
selection effect that favors detection of small orbits.  Jupiters
orbiting between 0.5 and 1.5 AU would be easily detected with our
current Doppler precision of 5 \ms, but none has been found.  This
distribution of orbits supports models in which orbital migration in a
gaseous protoplanetary disk drags jupiter--mass planets inward (Lin et
al. 1995, Trilling et al. 1998).

The distribution of the masses of substellar companions reveals two
populations.  Our survey of 107 GK dwarfs revealed none that had \msini
= 10 -- 80 \mjup $~$(Marcy \& Butler 1998).  Thus, ``brown dwarf''
companions occur with a frequency less than $\sim$1\%, within 5 AU.
Similarly, Mayor et al. (1997, 1999) surveyed $\sim$500 GK dwarfs, and found
at most 4 companions between 10 -- 80 \mjup. (Hipparcos astrometry
has shown that seven previously suspected brown dwarfs from that sample 
are actually H--burning stars.)  
In contrast, at least 5\% of GK stars harbor companions with
masses from 0.5 -- 5 \mjup.  For example, in our Doppler survey of 107
main sequence stars at Lick Observatory, we found 6 companions that have
$M\sin i $ = 0.5 -- 5 \mjup $~$(Marcy and Butler 1998, this paper).
Thus, relative to this well--populated planetary decade of
masses, there exists a brown dwarf ``desert'' at masses 10 -- 80 \mjup,
within 5 AU.

The efforts described above have focussed on G-- and K--type main
sequence stars having masses between 0.8 and 1.2 \msun.  The question
arises regarding the prevalence of planets around the M dwarfs which
constitute 70\% of the stars in the Galaxy.  Here we describe the
detection of the first apparent planetary companion to an M dwarf, 
Gliese 876, located 4.7 pc from the Sun.

\section{Observations}
\label{obs}

Gliese 876 (=HIP 113020) has V magnitude of 10.1, a spectral type of
M4V and a parallax from Hipparcos of 0.213 (Perryman et al. 1997).
Adopting this parallax and the bolometric correction of Delfosse et
al. (1998) gives $M_{\rm Bol}$ = 9.52, which implies a luminosity of,
$L$ = 0.0124 L$_{\odot}$.  The mass of the star Gliese 876 can be
derived from its K-band apparent magnitude (K=5.04) and parallax,
along with an empirical mass--luminosity relation (Henry \& McCarthy
1993). This gives $M_*$ = 0.32 $\pm$ 0.03 M$_{\odot}$.  Gliese 876 is
chromospherically inactive (Delfosse et al. 1998), which suggests that
it is older than $\sim$1 Gyr.  However its space motion is slow which
suggests that its age is less than 10 Gyr.  Its metalicity is not
known well, though a preliminary synthesis of the spectrum indicates
that it is metal poor by a factor of 2--3 relative to the Sun
(Valenti, 1998).

Doppler shifts for Gliese 876 have been obtained at both Lick and Keck
Observatories, using the Hamilton and HIRES echelle spectrometers,
respectively (Vogt 1987, Vogt et al. 1994).  The first observations
were made in 1994.9 (at Lick) and in 1997.4 (at Keck), and both data
sets extend to the present.  The calibration of wavelength and the
measurement of the spectrometer PSF was determined for each exposure
and for each 2--{\AA} chunk of spectrum by using iodine absorption
lines superimposed on the stellar spectrum (Butler et al. 1996).
Figures 1 and 2 show all of the individual velocity measurements as a
function of time, along with the separate Keplerian fits.  

The velocities from Lick Observatory have typical uncertainties of 30 \ms and
those from Keck are 6 m$~$s$^{-1}$.  Poisson statistics of the
photons dominate the velocity errors for this relatively faint
(V=10.1) star.  Error bars on all points are the uncertainty in the
mean of the velocities ($\sigma/\sqrt{N_{\rm chunk}}$) from the many
2--{\AA} wide chunks into which the spectrum was divided.  Doppler
measurements of Gliese 876 at Haute Provence by Delfosse et al. (Mayor
et al. 1999) also show an amplitude and periodicity in agreement with
those reported here, thus constituting an immediate confirmation.  It
remains to be seen if their orbital parameters agree with those quoted
here.

The Lick and Keck data each carry independent and arbitrary velocity
zero-points.  The relative zero--point 
has been determined by combining the two data sets and
adjusting the velocity offset until the Keplerian
fit (see \S 3) yields a minimum in the $\chi^2$ statistic .  
Thus, the Lick and Keck velocities 
were forced to have the same zero-point.

\section{Orbital Solution}

Independent Keplerian fits were determined from the Lick and Keck data
sets, and the resulting curves and orbital parameters are shown in
Figures 1 and 2.  The final orbital parameters are given in
Table~\ref{orbit}, based on an orbital fit to the combined data set.
The uncertainties reflect the differences in the two independent
orbital fits.  The two solutions agree within their uncertainties.
The joint orbital period is $P$ = 60.85 $\pm$ 0.15 d, and the
eccentricity is $e$ = 0.27 $\pm$ 0.03.  The orbital solution implies a
planetary orbital semi-major axis of 0.21 $\pm$ 0.01 AU, and a minimum
mass of \msini = 2.1 $\pm$ 0.2 \mjup.
This inferred \msini is proportional to the assumed mass
of the host star (0.32 $\pm$0.03 \msun) which contributes
most of the uncertainty in the companion mass.

The periodic repetition of an asymmetric radial velocity variation is
apparent from the raw data and from the fits in Figures 1 and 2.  The
orbit is clearly not circular.  There is no pattern in the residuals,
thus excluding the presence of any second planet with a mass greater
than 1 Jupiter mass and a period of 4 years or less in the Gliese 876
system.  The Lick and Keck velocities can be merged to yield a final
fit, as shown in Figure 3.  This shows that the two sets share a
common orbital phase in addition to similar best--fit orbital
parameters.  We note that two points from Lick sit off the Keplerian
curve by 2$\sigma$, and we suspect that the quoted errors of $\sim$30
\ms in those cases may be underestimated due to the low
signal--to--noise ratios of those spectra.  


The large velocity amplitude of 220 \ms for Gliese 876 leaves orbital
motion as the probable cause of the velocity variations.  Spots on a
rotating star can, in principle, cause artifical velocity variations.
But for Gliese 876, the equatorial rotation velocity is less than 2
km$~$s$^{-1}$, and the star is photometrically stable to within
$\sim$0.02 mag (Marcy and Chen 1992, Weiss 1996, Delfosse et
al. 1998).  Therefore, spots cannot alter the apparent velocity by
more than $\sim$0.02 $\times$ 2000 \ms = 40 \ms .  We have not checked
for stellar pulsations, but the photometric stability suggests that
any pulsations are not significant here.  Moreover, acoustic
oscillations and g--modes for a 0.3 M$_{\odot}$ dwarf would have time
scales of minutes and hours, respectively, unlike the observed 60 day
velocity period.

\section{Discussion}

The companion to Gliese 876, with \msini = 2.1 $\pm$0.2 \mjup, has
a likely mass of 2 to 4 \mjup, assuming unbiased orbital inclinations.
For an assumed companion mass of 2.1 \mjup, the astrometric semimajor
axis would be 0.28 mas.  Hipparcos astrometry exhibits no wobble at a
2--$\sigma$ upper limit of 4 mas (Perryman et al. 1997).  Thus, the
upper limit to the companion mass is 29 \mjup.

At 4.7 pc, this is the closet known extrasolar planet.  The semimajor
axis implies an angular separation 0.045 arcsec, with a greatest
separation of 0.062 arcsec.  It is thus a prime candidate for direct
imaging with IR adaptive optics and with interferometry (i.e., Keck,
LBT, SIM, VLTI).  Astrometric detection is also favored due to:  1. its
close proximity to the Sun, 2. the large mass of the planet, 3. the
low mass of the star, 4. the small orbital period which permits many
cycles to be monitored within a season.

Gliese 876 is apparently the first M dwarf with a known planetary
companion.  We have surveyed only 24 M dwarfs from Lick Observatory
during the past 4 yr (with poor precision of 25 \ms), which implies
that the occurrence of Jupiter--mass planets within 2 AU of M dwarfs
could be a few percent, based on this one detection.  The duration and
paucity of Keck observations render them not yet adequate ($\sim$1 yr)
to add information on the occurrence of planets around M dwarfs.

The small orbital semi-major axis of $a$=0.21 AU and the eccentricity of
$e$=0.27 pose two profound puzzles regarding the origin of such
planetary orbits.  There is too little mass within a planetary feeding
zone in a nominal protoplanetary disk at distances of 0.2 AU to
provide 2 Jupiter--masses of material to a growing planet
(cf. Lissauer 1995).  One suggestion is that giant planets form
several AU from the star and then migrate inwards.  Orbital migration
can be induced by interactions between the planet and the gas in the
protoplanetary disk, bringing the planet inwards (Lin et al. 1995,
Trilling {\it et~al.\/} 1998).

However, it is not clear what would cause the planet around Gliese 876
to cease its migration at 0.2 AU.  Neither tidal interactions with the
star nor a magnetospherically--cleared hole at the disk--center would
extend to 0.2 AU, and thus they cannot halt the migration.  A similar,
as--yet-unidentified parking mechanism appears needed for the planets
around 55 Cancri and $\rho$ Cor Bor (Noyes et al. 1997, Butler et al.
1997).

The non--circular orbits for both $\rho$ Cor Bor ($e$=0.16 $\pm$ 0.06)
and for this planet around Gliese 876 ($e$=0.27$\pm$0.03) imply that
significant orbital eccentricities are common for Jupiter--mass
companions orbiting between 0.1 and 0.3 AU from their star.  Some
physical mechanism must be identified which generally produces sizable
eccentricities, in contrast to the inexplicably low eccentricities of
the Giant Planets in our Solar System.  Infrared speckle reveals no
companions to Gliese 876 from 1 AU outward (Henry \& McCarthy 1990),
and the lack of large variations in the velocities rule out stellar
companions within 1 AU.  Thus, the eccentricity of the planetary
companion around Gliese 876 could not have been pumped by a stellar
companion.

Apparently, migration, if necessary, did not enforce circularity in
the final orbits of Gliese 876 or $\rho$ Cor Bor.  
One possible explanation is that gravitational
scattering of planetary cores (of Earth--mass and larger) can dominate
the orbital evolution (Rasio and Ford 1996, Weidenschilling and
Marzari 1997, Lin and Ida 1996).  Orbit--crossings and global
instabilities among planetesimals in the disk can lead to dramatic
orbit changes and large eccentricities (Levison et al. 1998).

Long--lived gas in a protoplanetary disk may lead to circular orbits
in such planetary systems.  Other systems that lose their gas may
suffer dynamical instabilities, leading to eccentric orbits at a
variety of semimajor axes.  However, the latter scenario, if common, 
does not explain the apparent paucity of jupiters from 0.5 to 1.5 AU, and it
remains to be seen if jupiters are common farther out.

The equilibrium temperature at optical depth unity in the atmosphere
of the planet around Gliese 876 is estimated to be -70 C, too cold for
water in liquid form (Saumon 1998).  Temperatures would be higher at
deeper layers in the atmosphere.  Any bodies orbiting interior to 0.2
AU would have surface temperatures above -70 C.  It would be
interesting to determine if planets could reside in stable orbits
within 0.2 AU, perhaps in mean--motion resonances with the giant
planet discovered here.

\acknowledgements

We thank Kevin Apps for analysis of Hipparcos astrometry.  We thank
Xavier Delfosse, Michel Mayor, and Didier Queloz for communicating
their velocities for Gliese 876.  We thank M. Duncan, D. Lin, and
G.Basri for useful conversations.  We acknowedge support by NASA grant
NAGW-3182 and NSF grant AST95-20443 (to GWM), and by NSF grant
AST-9619418 and NASA grant NAG5-4445 (to SSV) and by Sun Microsystems.
We thank the NASA and UC Telescope assignment committees for
allocations of telescope time.

\clearpage
\begin{figure}
\figcaption{The Lick radial velocities for Gliese 876 obtained
from 1994 to 1998.6 vs orbital phase.  
The solid line is the radial velocity curve from the best--fit 
orbital solution from the Lick data alone.}
\label{rv_curve}
\end{figure}

\begin{figure}
\figcaption{The Keck radial velocities for Gliese 876.
The solid line is the radial velocity curve from the orbital solution
from the Keck data alone.}

\label{fig2}
\end{figure}

\begin{figure}
\figcaption{The combined Lick and Keck radial velocities for Gliese 876,
plotted versus orbital phase.  Filled circles are from Lick and triangles
come from Keck.
The solid line is the radial velocity curve from the orbital solution.}
\label{fig3}
\end{figure}

\clearpage

\begin{deluxetable}{lcc}
\tablecaption{Combined Orbital Solution for Gliese 876
\label{orbit}}
\tablewidth{0pt}
\tablehead{
\colhead{Parameter} & \colhead{Value} & \colhead{Uncertainty}
}
\startdata
Orbital Period $P$ (days)                 &  60.85 & 0.15 \\
Velocity Semi-amplitude $K$ (m\,s$^{-1}$) & 239  & 5 \\
Eccentricity $e$                          & 0.27 & 0.03 \\
Longitude of Periastron $\omega$ (deg)    & 24   & 6 \\
Periastron Date $T_0$ (Julian Date)       & 2450301.0 & 1.0 \\
$M \sin i$ ({\rm M}$_{\rm JUP}$)          & 2.11 & 0.20   \\
Semimajor Axis a (AU)                     & 0.21 & 0.01   \\ 
\enddata
\end{deluxetable}
 
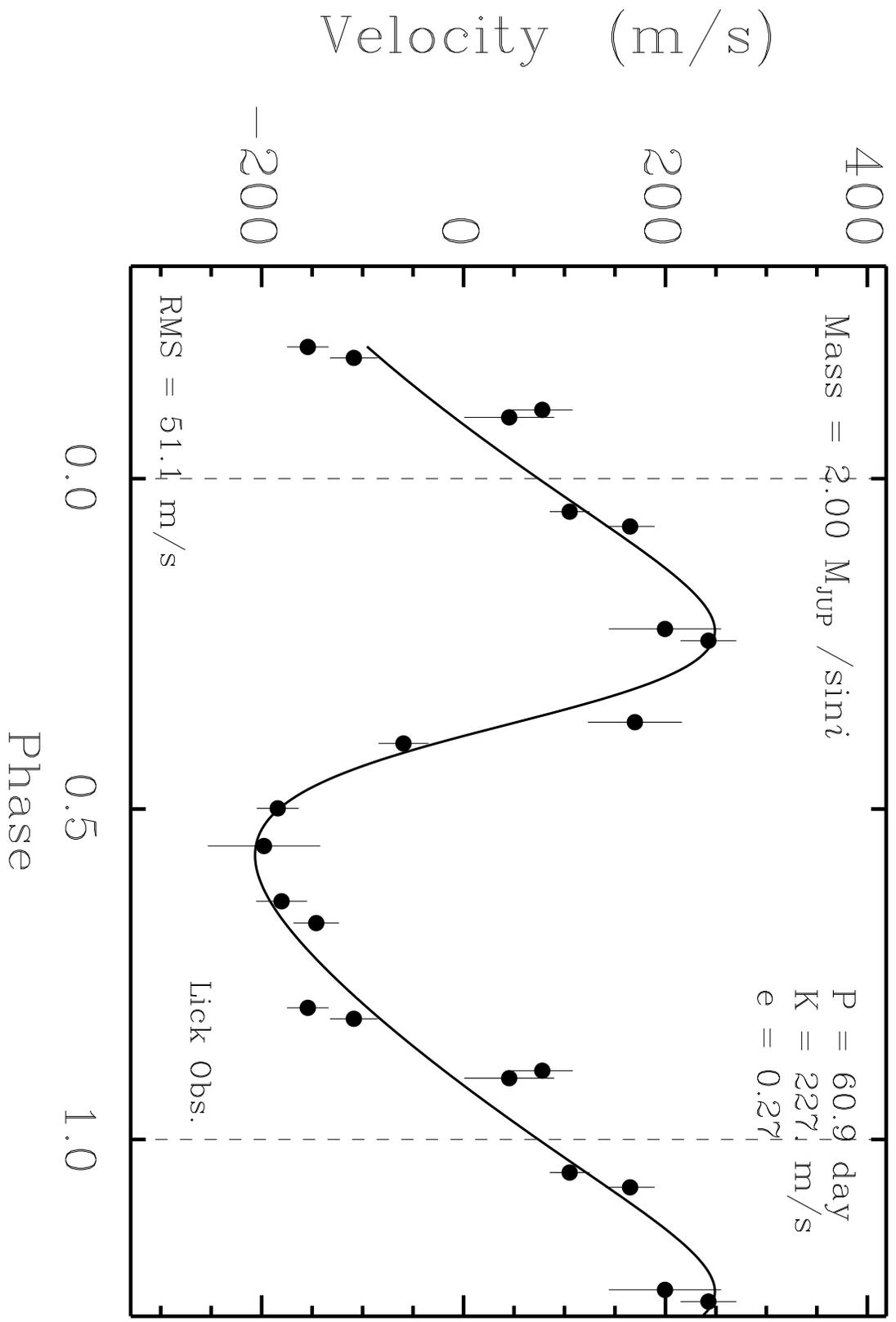
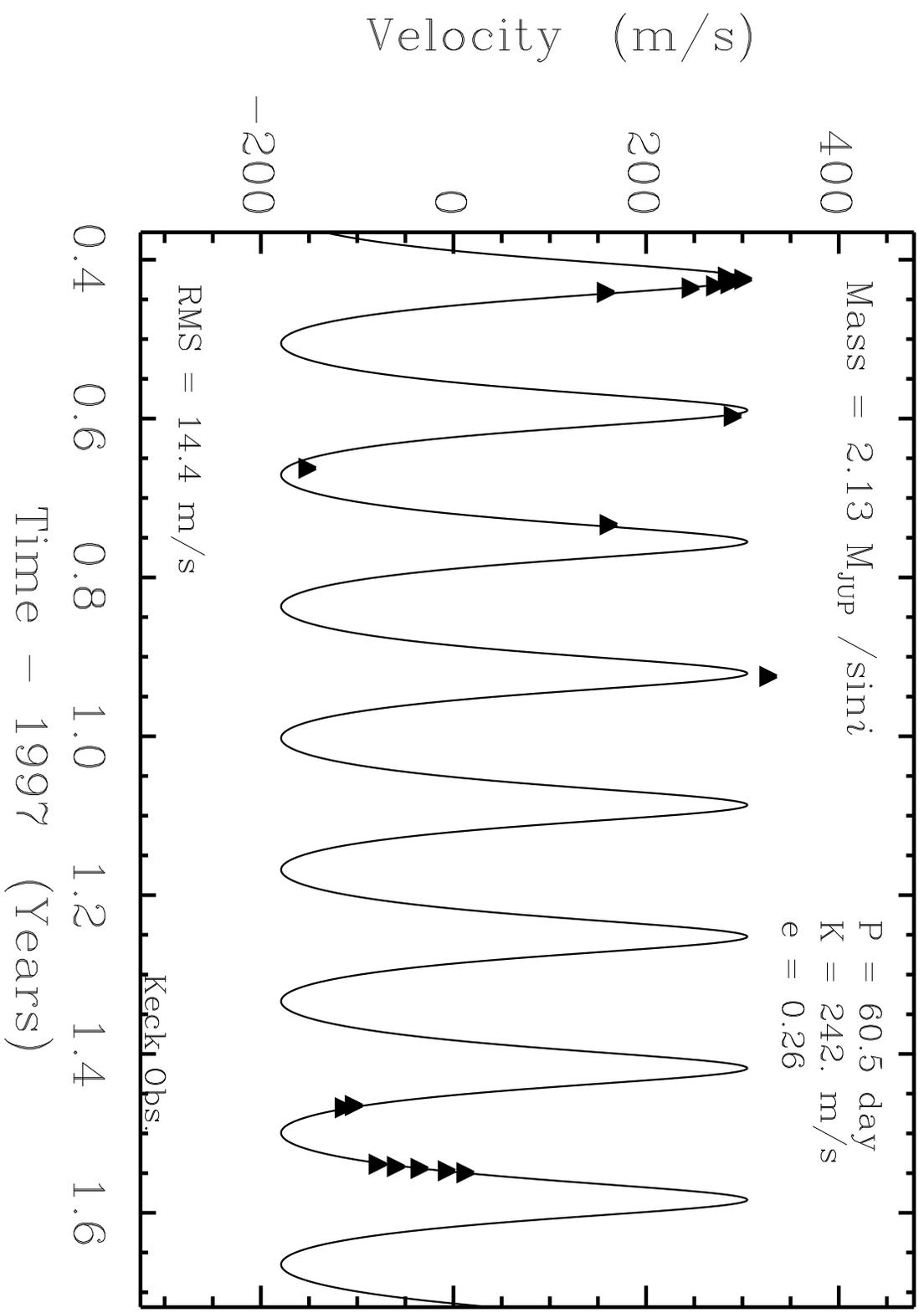
\end{document}